\documentclass[useAMS,usenatbib]{mn2e}
\usepackage{graphicx}
\usepackage{amssymb}
\title[On the rotational behavior of parent stars of extrasolar planets]
{On the rotational behavior of parent stars of extrasolar planets
\thanks{}
}
\author[S.  Alves]{S. Alves$^{1}$ \thanks{E-mail: sanzia@dfte.ufrn.br},
 J.D. do Nascimento Jr$^1$\thanks{E-mail: dias@dfte.ufrn.br},
 J.R. De Medeiros$^1$\thanks{E-mail: renan@dfte.ufrn.br}\\
$^1$ Departamento de F\'{\i}sica, Universidade
Federal do Rio Grande do Norte, 59072-970, Natal, RN.,
Brazil}

\newcommand{\Msun}{${M}_\odot$}

\newcommand{\Teff}{${T}_{\textrm{eff}}$\ }

\begin{document}

\date{Accepted for publication in the MNRAS }

\pagerange{\pageref{firstpage}--\pageref{lastpage}} \pubyear{2010}

\maketitle

\label{firstpage}

\begin{abstract}
We analyzed the behavior of the rotational velocity in  the parent stars of extrasolar planets. Projected rotational velocity $v \sin i$ and angular momentum were  combined with stellar and planetary parameters, for a unique sample of 147 stars, amounting to 184 extrasolar planets, including 25 multiple systems. Indeed, for the present working sample we considered only  stars with planets detected by the radial-velocity procedure.  Our analysis shows that the $v \sin i$ distribution of stars with planets along the HR Diagram follows the well established scenario for the rotation of intermediate to low main sequence stars, with a sudden decline in rotation near 1.2~\Msun. The decline occurs around \Teff $\sim$ 6000~K,  corresponding to the late-F spectral region. A statistical comparison of the distribution of the rotation of stars with planets and  a sample of stars without planets indicates that the $v \sin i$ distribution for these two families of stars is  drawn from the same population distribution function. We also found that the angular momentum of extrasolar planet parent stars follows, at least qualitatively, Kraft’s relation $ J \propto (M/M_{\odot})^\alpha$. The stars without detected planets show a clear trend of angular momentum deficit compared to the stars with planets, in particular for masses higher than about 1.25~\Msun. Stars with the largest mass planets tend to have angular momentum comparable to or higher than the Sun.
\end{abstract}


\section{Introduction}
Since the pioneering discovery by Mayor \& Queloz (1995) of a Jupiter-mass planet orbiting 51 Peg star , a large number of extrasolar planets have been discovered orbiting stars of different evolutionary stages and masses. As of July 30th, 2008, there were 322 planetary-mass companions known to be orbiting mainly  solar-type stars, including 35 multi-planet systems. During the past decade these discoveries have inspired intensive studies on the physical properties of the planets and of their parent stars. The discovered planets have masses ranging from 4 Earth masses to 11 Jupiter masses.  They can be found at distances of several AU or close to the parent star, with orbital periods in the range of a few days to a few years. High eccentricity is a common parameter connecting these planets (e.g.: Marcy et al. 2001).

The nearby dwarf stars with giant planets show evidence of moderate metal enrichment compared to the average metallicity of field dwarfs in the solar neighborhood without detected planets. The dependence of planetary frequency on the metallicity of  host dwarf stars has been investigated since they were first detected  by precision radial-velocity surveys (Gonzalez 1997; Laugh-lin \& Adams 1997). Different explanations have been proposed for this dependence, such as enhanced giant planet formation by high stellar metallicity (Santos et al. 2000,2001; Reid 2002), observational selection effects or pollution by ingested planetary material (Laughlin 2000; Gonzalez et al. 2001; Israelian et al. 2001; Pinsonneault et al. 2001; Murray \& Chaboyer 2002). For instance, based on observationally unbiased stellar samples, the evidence for higher planetary frequency around unpolluted, primordially metal-rich stars has been clearly demonstrated by Santos et al. (2001), and confirmed by Fischer et al. (2003) and Santos et al. (2004), who showed a sharp break in frequency at [Fe/H]~$\sim$~0.0. Further, several studies have investigated a possible link between the orbital period of an extrasolar planet sample and the metallicity of the parent stars. Some authors have argued in favor of this hypothesis  (Gonzalez 1998; Queloz et al. 2000; Jones 2003), while others (Santos et al. 2001; Laws et al. 2003) found no evidence of its existence.  In particular, Santos et al. (2003) concluded that the metallicity distribution of stars with very short period planets, typically $P_{\rm orb} < 10$ days, is essentially indistinguishable from the same distribution of stars with longer orbital period planets, whereas Sozzetti (2004) claims there is a possible correlation between stellar metallicity and planet orbital period, where close-in planets, in an orbit of a few days, would more likely be found around metal-rich stars.  In contrast to the behavior of metallicity in dwarf stars with planets, giant stars hosting planets do not show a tendency to being more metal rich (Pasquini et al. 2007). Mazeh et al. (2005) pointed out an intriguing correlation between the masses and orbital periods of extrasolar planets, consistent with a linear relation, at least for planets with periods shorter than five days. Such a result was confirmed more recently by Torres et al. (2008). In addition, Gonzalez (2008) confirmed that, near the solar temperature, Li abundances of stars with planets are smaller than those of stars without planets. For a solid review, presenting a census of the main statistical results obtained to date in this domain and a discussion on extrasolar planet general orbital properties, the reader is referred to Udry and Santos (2007).

Among other possible defining characteristics of stars with planets, little attention has been paid to the analysis of their rotational behavior.  More specifically, a comprehensive study on the rotational behavior of stars harboring extrasolar planets, in spite of the obvious importance of rotation in stellar and planetary system evolution, has yet to be performed. Actually, the vast majority of recently discovered planetary systems contain solar-type stars and this fact suggests  that stars with planets are  slow rotators at their respective ages. Indeed, only a preliminary study by Barnes (2001), based on the rotation periods of 35 such stars, focused on  this subject, , suggesting that planet host stars have normal rotational properties, whereas a more recent study by Gonzalez (2008) points to a scenario where the $v \sin i$ values of stars with planets are smaller than those of stars without planets  for stars cooler than 6000~K, nearly the same for stars near 6000~K and much larger for the hottest stars.

This work presents an unprecedented analysis of the behavior of the projected rotational velocity $v \sin i$ of stars with planets, using the current sample of such stars available in the literature, in addition to searching  for a possible connection between stellar rotation and planetary parameters. An analysis of the angular momentum behavior of these stars is also performed. To make headway in the study of the rotational behavior of stars with planets,  it is, therefore, essential to conduct  a comparative analysis of the rotation and angular momentum behavior of stars without detected planets.  This is one of the major goals of the current study. The paper is organized as follows: in Section 2 we present the characteristics of the working sample and in Section 3 we analyze possible sources of bias that might contribute to producing  the features observed.  In Section 4 we present our findings, with a brief discussion and finally, conclusions are outlined in Section 5.

\section{Stellar working sample}

This work is based on the sample of stars with extra-planets listed in the comprehensive Extrasolar Planets Encyclopedia, maintained by J. Schneider, as of July 30th, 2008,  amounting to 322 planetary-mass companions known to be orbiting 272 almost solar-type stars, including 35 multiple planet systems. Nevertheless, in the final working sample we  considered only the stars with planets detected by radial-velocity procedure, consisting of  147 stars with projected rotational velocity $v \sin i$ available in the aforementioned Encyclopedia, hosting 184 planets, including 25 multiple systems. It should be pointed out  that all these stars are listed in the Geneva planetary search survey.  With such a  criterion, all the stars with transiting planets, selected only from photometric surveys, are purged. This fact automatically eliminates a possible bias favoring fast rotators coming from transiting systems. The main stellar and planetary physical observables used in this study, namely, $v \sin i$, mass and stellar age, as well as the planetary orbital parameters, can be retrieved from Schneider (2009) and references therein. Readers are referred to these references for a discussion of measurement procedures and error analyses of the stellar parameters listed.
\begin{figure}
\centering
\resizebox{1\hsize}{!}{\includegraphics{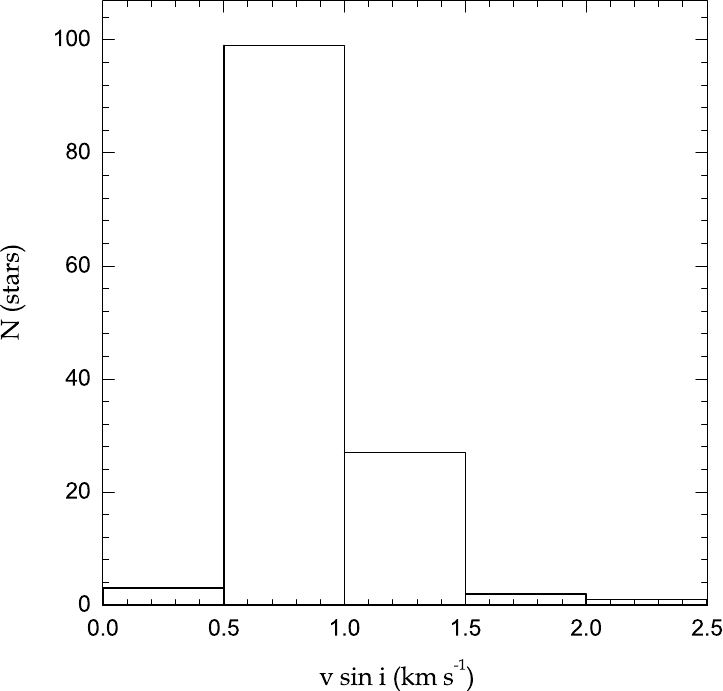}}
\caption[]{The distribution of errors of rotational velocity $v \sin i$ measurements for the  sample of 147 parent stars of extrasolar planets used in the present study.}
\label{fig1.ps}
\end{figure}

For comparative purposes, we took two samples of stars not known to have any planetary-mass companions, one  from the Geneva planet search survey composed of 39 F- and G-type dwarf stars (Santos et al. 2004,2005) and the other from the Anglo-Australian planet survey composed of 85 dwarf stars (Bond et al. 2008). However, we should be cautious about these samples, which derive from a list of stars that are surveyed for planets, but for which none have yet been found.  Certainly, this does not mean that such stars  have  no planetary companions whatsoever. For instance, they might host planets with very low mass and/or long orbital period that are more difficult to detect with radial velocity surveys.

Because projected rotational velocity  $v \sin i$  is perhaps the most sensitive parameter in the present study, let us briefly discuss the quality and precision of the data obtained. For stars with planets,  $v \sin i$  measurements were taken from Valenti \& Fischer (2005) (91 stars), Holmberg et al. (2007) (26 stars), Marcy \& Chen (1992) (4 stars), da Silva (2006, 2007) (4 stars), Fischer et al. (2007) (3 stars), Bakos et al. (2007) (2 stars), Lovis et al. (2005) (2 stars), Santos et al. (2007) (2 stars), Tamuz et al. (2008) (2 stars) and 1 star each from Bonfils et al. (2007), Burke et al. (2007), Ge et al. (2006), Laughlin (2005), Mayor et al. (2009), McCullough et al. (2006), Naef et al. (2007), O’Donavan (2006), Win et al. (2007), Johnson (2006) and Wright et al. (2007).  Measurements from Holmberg et al. (2007) were computed using the {\it CORAVEL} cross-correlation method, with typical errors of about 1 km~s$^{-1}$, whereas those from Valenti \& Fischer (2005) were computed using the spectral synthesis procedure, with errors of about 0.5 km~s$^{-1}$. Figure~1 presents the error distribution in $v \sin i$, ranging from about 0.5 to about 2.5 km~s$^{-1}$, with a mean error of 0.6$\pm$ 0.3 km~s$^{-1}$.

For both stellar samples, the stellar effective temperature was computed from the \Teff--$(B-V)$ calibration given by Flower (1996), whereas the stellar radius was estimated according to Lang (1980).  Stellar luminosity $L$ was estimated following the same procedure employed by do Nascimento et al. (2000), using the parallaxes and the $m_{\rm V}$ magnitudes given by {\it HIPPARCOS} to derive the intrinsic absolute magnitudes $M_{\rm V}$. { {Stellar age, all computed on the basis of isochrones  interpolation (e.g.:  Saffe et al. 2005; Holmberg et al. 2009; da Silva et al. 2007) were taken from Schneider (2009) for stars with planets, and from Fischer and Valenti (2005) for stars without planets, whereas masses, all determined spectroscopically, were taken from Schneider (2009) and Valenti and Fischer (2005), respectively for stars with and without planets.}}

We also analyzed the behavior of stellar angular momentum, which was computed by assuming stars are a solid and uniform density sphere.  For the calculation of this parameter we used the relation for the mean stellar angular momentum $J(M)$, given by
\begin{equation}
    J(M)=\frac{\it {v}\sin \it{i}}{R(M)}I(M),
\end{equation}
where $I(M)=\frac{2}{5}MR^2$ is the solid body moment of inertia for a sphere, $v \sin i$ iis the projected stellar rotational velocity and $R(M)$ the radius of the star.

\subsection{Metallicity bias and $v \sin i$ of stars with planets}

As stated by Gonzalez (2003), three main sources of bias have been identified as being potentially important to the study of the physical properties of stars with planets: (i) Biases linked to specific determination of the metal content of the host stars.  As pointed out by Laws et al. (2003) and Gonzalez et al. (2003), spectroscopic determination of [Fe/H] seems to be more reliable than photometric methods. (ii) Biases in statistical analyses because of the sample limitation of stars with [Fe/H]~$< 0.0$. (iii) Biases linked to intrinsic metal line behavior. High-precision radial velocity in metal- rich surveys are in principle easier than in metal-poor ones.
\begin{figure}
\centering
\resizebox{1\hsize}{!}{\includegraphics{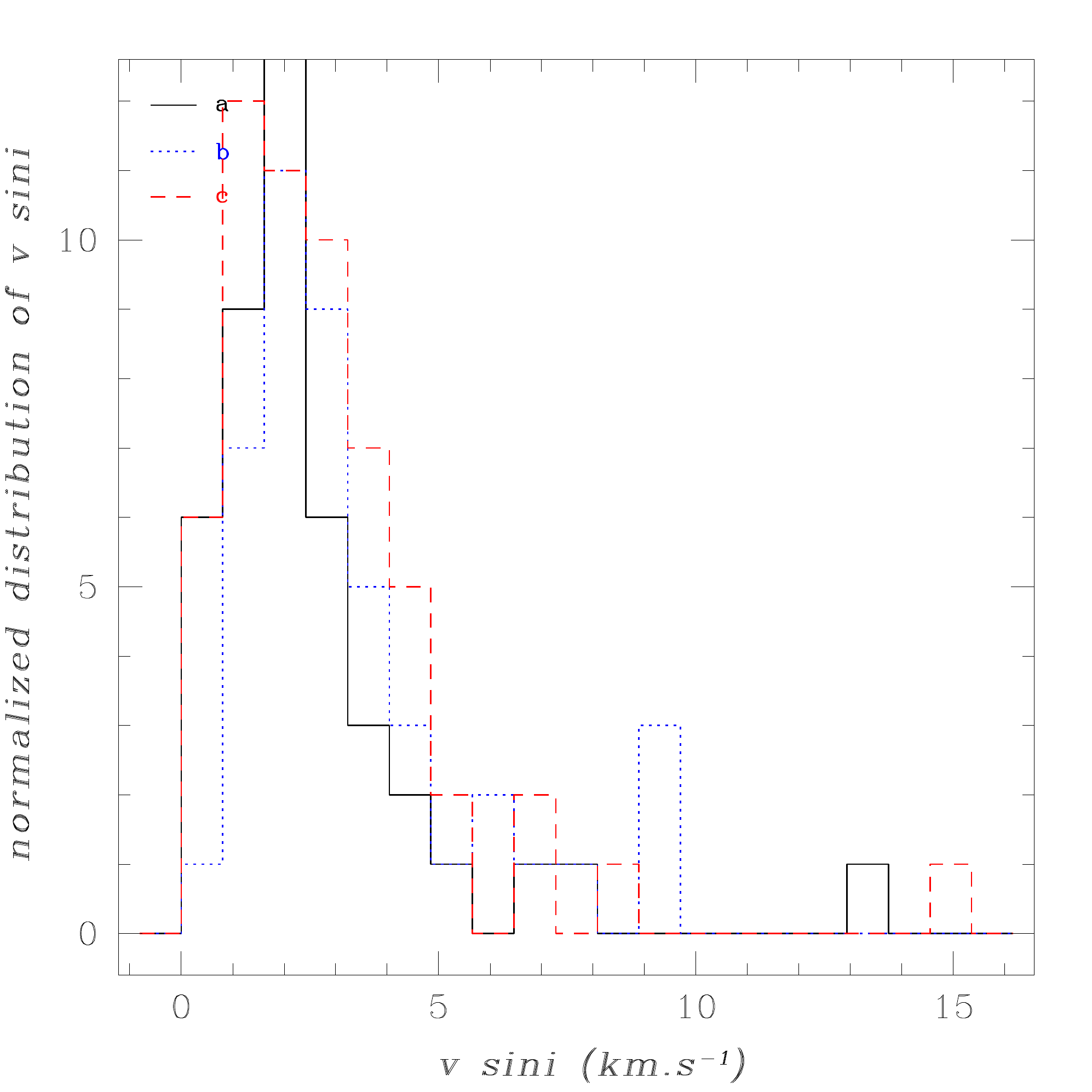}}
\caption[]{The normalized superimposed $v \sin i$ distribution of stars in the present sample. We divided the sample into three bins, (a) $-$0.71~$\leq$~[Fe/H] ~$<$~ 0.00 slightly underabundant in [Fe/H], (b) 0.00~ $\leq$ ~[Fe/H] ~$<$~ 0.21 around solar abundance, and (c) 0.21~$\leq$~ [Fe/H] ~$<$~ 0.50 slightly overabundant in [Fe/H].}
\label{fig2.ps}
\end{figure}

The fundamental method used to analyze the influence of these biases comes from Santos et al. (2003) and Fischer et al. (2003). By calculating the median velocity error as a function of metallicity for the stars in their planet  surveys, these authors found a velocity degradation of up to 50 per cent for the lowest metallicity stars ($[Fe/H] \sim-0.5$). Radial velocity surveys currently attain typical single measurement precision $\sigma_{RV} \sim~3-5$ m \thinspace s$^{-1}$, indicating that planets would be easily detected even with $\sigma_{RV} \sim~ 5-8 $ m \thinspace s$^{-1}$. One can therefore conclude that radial velocity precision degradation for metal- poor stars is not the major cause of any correlation.  To explain the fact that the lower occurrence rate of stars with planets around metal -poor stars had  little influence on our results, we divided the working sample of 147 stars into three bins, (a) $-0.71~\leq$~[Fe/H] ~$< 0.00$ slightly underabundant in [Fe/H], (b) $ 0.00~\leq$~[Fe/H]~$< 0.21$ around solar abundance, and (c) $0.21~\leq$~[Fe/H]~$< 0.50$ slightly overabundant in [Fe/H]. For this analysis we used only spectroscopic determinations of [Fe/H]. The distributions for the three different bins are displayed in Fig. ~2, showing no clear trend for $v \sin i$ in the metallicity ranges considered.

\section{Results and Discussion}

The aim of this pioneering study is to initiate a fruitful discussion to make the community aware of the fundamental role of rotation in our understanding of extrasolar planets and their relation with central stars.  To this end, we dedicated most of our efforts to identifying qualitative trends between rotation and different stellar and planetary parameters. 

\subsection{The rotation of stars with planets}

The distributions of the projected rotational velocity $v \sin i$ in the HR Diagram for our  working samples of 147 stars with planets and 85 stars without detected planets is displayed, respectively, in the upper and lower panels of Fig. 3. The Sun is also indicated for comparative purposes. Evolutionary tracks, with solar metallicity, follow the procedure used by do Nascimento et al. (2000). The first interesting feature emerging from this figure is the fact that the present sample of main sequence stars with planets, once displayed in the HR Diagram, tends to follow the well established scenario for the rotation of intermediate to low main-sequence stars, with a sudden decline in $v \sin i$ values near 1.2~\Msun. Such a scenario was first pointed out in Kraft’s (1976) remarkable paper on the strong observational basis emerging from $v \sin i$ measurements for late-F and early-G dwarfs in nearby young clusters and in the field.  This decline occurs around \Teff~$\sim$~6000~K, corresponding to the late-F spectral region (e.g.: Soderblom 1983). Indeed, a detailed inspection of the aforementioned  figure shows that the rotational behavior of main-sequence stars with planets parallels that  observed in the milestone HR Diagram presented by Kraft (1967), where stars with enhanced rotation are mostly those with effective temperature higher than about 5800~K and luminosity above the solar value, whereas slow rotators are mostly cooler stars, with luminosity lower than the solar value. The fast and moderate transition to slow rotators is explained as resulting from the presence of a convective envelope in late-type stars that slows them down from an initially rapid rotation rate through magnetic breaking, associated to the fact that these stars are on average older than the early-type and have undergone a longer  spin-down period. In spite of the smaller number of stars without detected planets, relative to the sample of stars with planets, a comparative inspection of panels displayed in Fig.~ 3 shows a clear trend for similar distributions of the projected rotational velocity $v \sin i$ for both families of stars, in the region of the HR Diagram defined by $0.6~\gtrsim~\log($L/L$_{\sun})\gtrsim~-0.2$ and $3.8~\gtrsim~\log$~\Teff~[K]~$~\gtrsim~3.7$.

\begin{figure}
\centering
\resizebox{1\hsize}{!}{\includegraphics{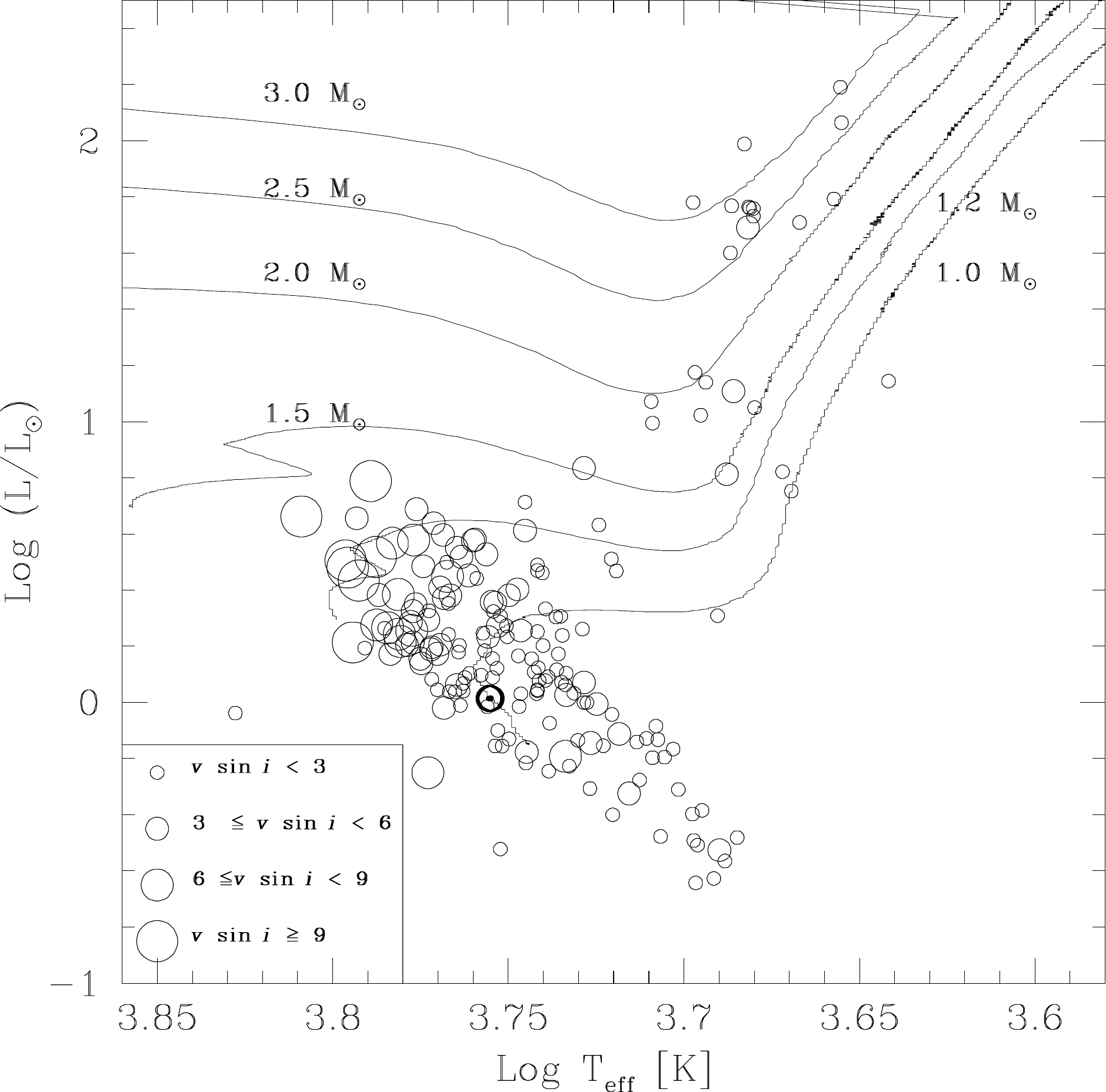}}\\
\resizebox{1\hsize}{!}{\includegraphics{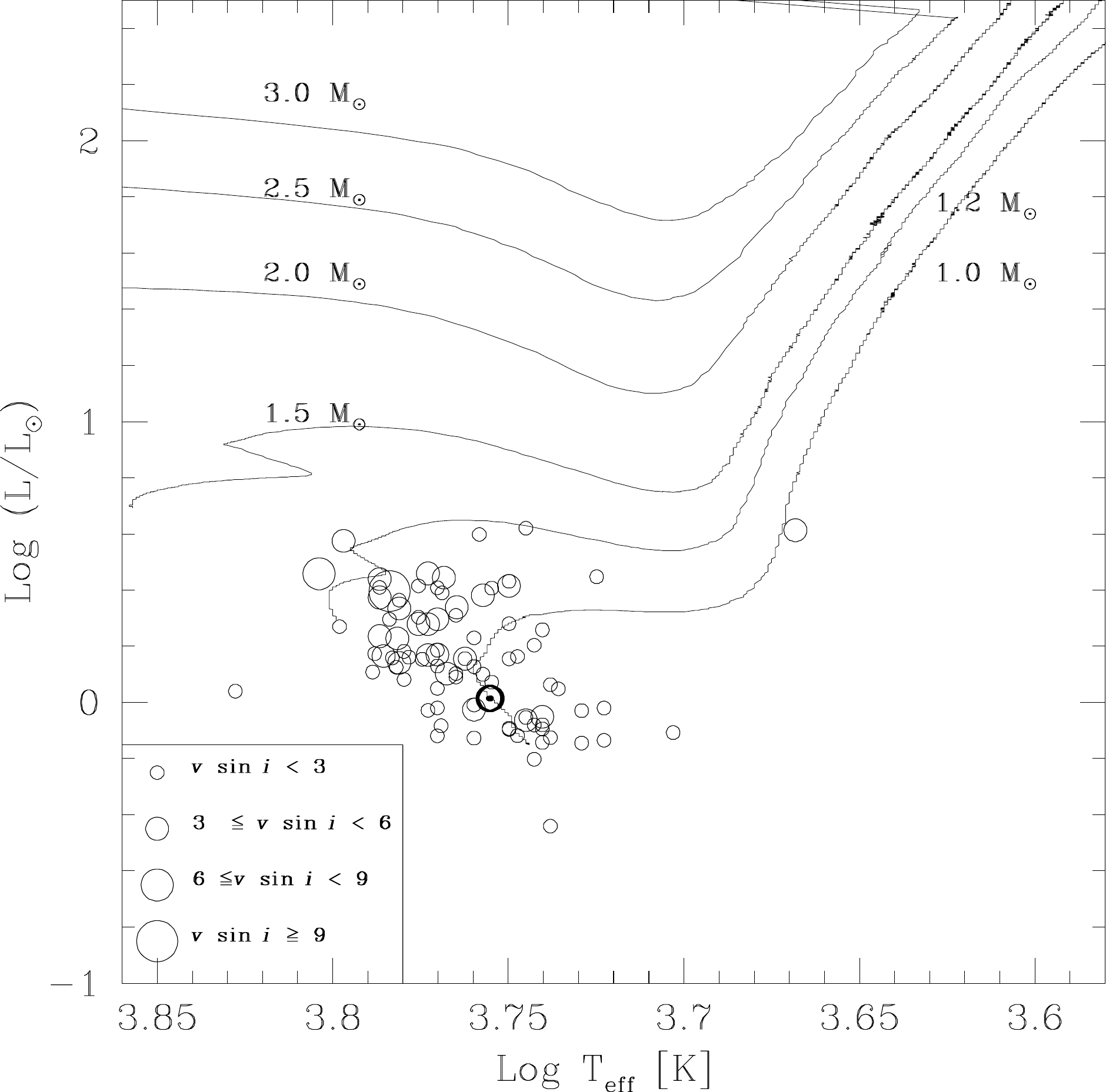}}
\caption[]{The distribution of rotational velocity $v \sin i$ for 147 parent stars of extrasolar planets (upper panel) and for 85 stars without detected planets from Bond et al. (2008) (lower panel). Evolutionary tracks for solar metallicity followed the procedure used by do Nascimento at al. (2000). The sun is also represented for comparative purposes.}
\label{fig3}
\end{figure}

In spite of the scenario observed in the distribution behavior of $v \sin i$ in  planet host stars along the HR-Diagram, while following the scenario discovered by Kraft (1967), one important question emerges at this point: Are the rotational properties of planet host stars normal compared to the stars without detected planets? Although objective comparisons between the parameters of stars with planets and those of stars without planets have proven to be difficult, the present study offers the possibility of a statistical comparison between the rotation distributions in these two families of stars.  Indeed, the selection criteria of the present radial-velocity planet search surveys include stars exhibiting sufficiently sharp lined spectra, to detect the small Doppler shifts induced by planetary mass companions and, in this context, both samples defined in the previous section exclude stars with very large $v \sin i$ values, exhibiting only slow to moderate rotators.

In order to verify if the present data sets are significantly different from one another, we performed a Kolmogorov-Smirnov test (Press et al. 1992), which calculates the probability that two distributions are derived from the same parent distribution.  According to the K-S test, zero probability means the distributions are dissimilar, whereas unit probability means they are the same.  During the present study we applied the Kuiper’s K-S statistic (e.g.: Jetsu \& Pelt 1996; Paltani 2004), hereafter the K-S test, which, in contrast to the standard Kolmogorov-Smirnov statistic, is invariant under an origin shift for distributions, offering the same weight for all the points of a distribution.

\begin{figure}
\centering
\resizebox{1\hsize}{!}{\includegraphics{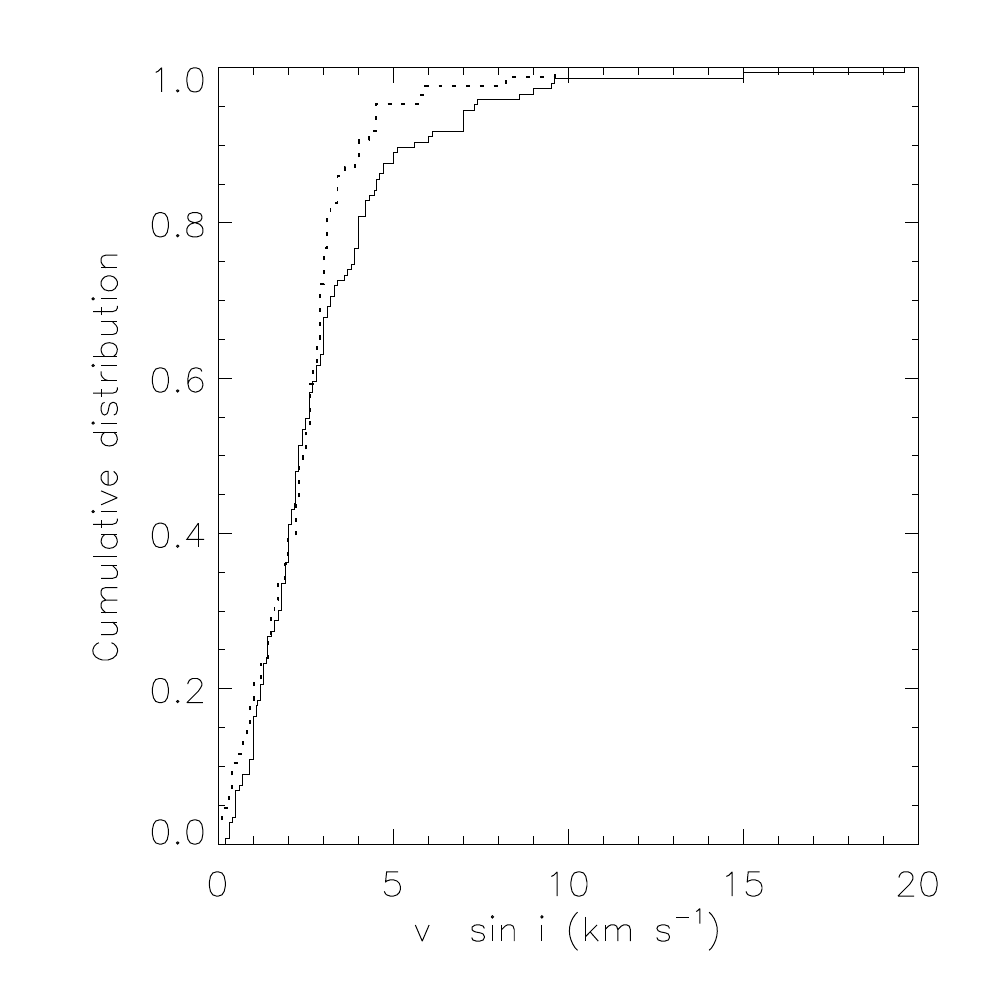}}
\caption[]{Comparison of the distribution of rotational velocity $v \sin i$ of main–sequence stars with and without planets. Solid and dashed curves stand for the cumulative distribution of $v \sin i$ of stars with and without planets, respectively.}
\label{fig4}
\end{figure}
As a first step, we  performed  K-S analyses, taking into account all the main-sequence stars listed in our working sample, namely 147 stars with planets and 85 stars without detected planets.  Fig. ~4 shows the cumulative functions for both $v \sin i$ distributions..The probability value  of 0.209  obtained on the K-S test is consistent with the two distributions being drawn from the same population.  In addition, we  performed the K-S analyses by comparing groups of stars that are  in the same region of the HR Diagram.  Samples of dwarf stars with and without planets were grouped first by spectral types, namely F- and G-type stars, and then by effective temperature, namely stars cooler and hotter than 6000 K. Now taking into account the sample of 118 F- and G-type main-sequence stars with planets, defined in the present work, and 82 F- and G-type stars without planets, listed by Bond (2008), a probability value of the K-S test of 0.299 is obtained, reinforcing the previous scenario with the two distributions of $v \sin i$ being drawn from the same parent population. In addition, if we compare similar samples of F- and G-type dwarf stars from the Geneva survey, a probability value of 0.947 is obtained, indicating that $v \sin i$  distributions are drawn from the same parent distribution, at a 93\% confidence level. 

The KS-test applied for main-sequence stars with planets segregated by effective temperature, namely cooler and hotter than 6000~K, gives a probability of $4 \times 10^{-6}$, pointing towards a scenario where the distributions of $v \sin i$ for  stars with planets, hotter and cooler than 6000 K are in fact not drawn from the same population distribution function. Such a result reinforces the scenario observed in Fig.~3, with a clear decline in $v \sin i$ of dwarf stars with planets around \Teff $\sim$ 6000~K.

\begin{figure}
\centering
\resizebox{1\hsize}{!}{\includegraphics{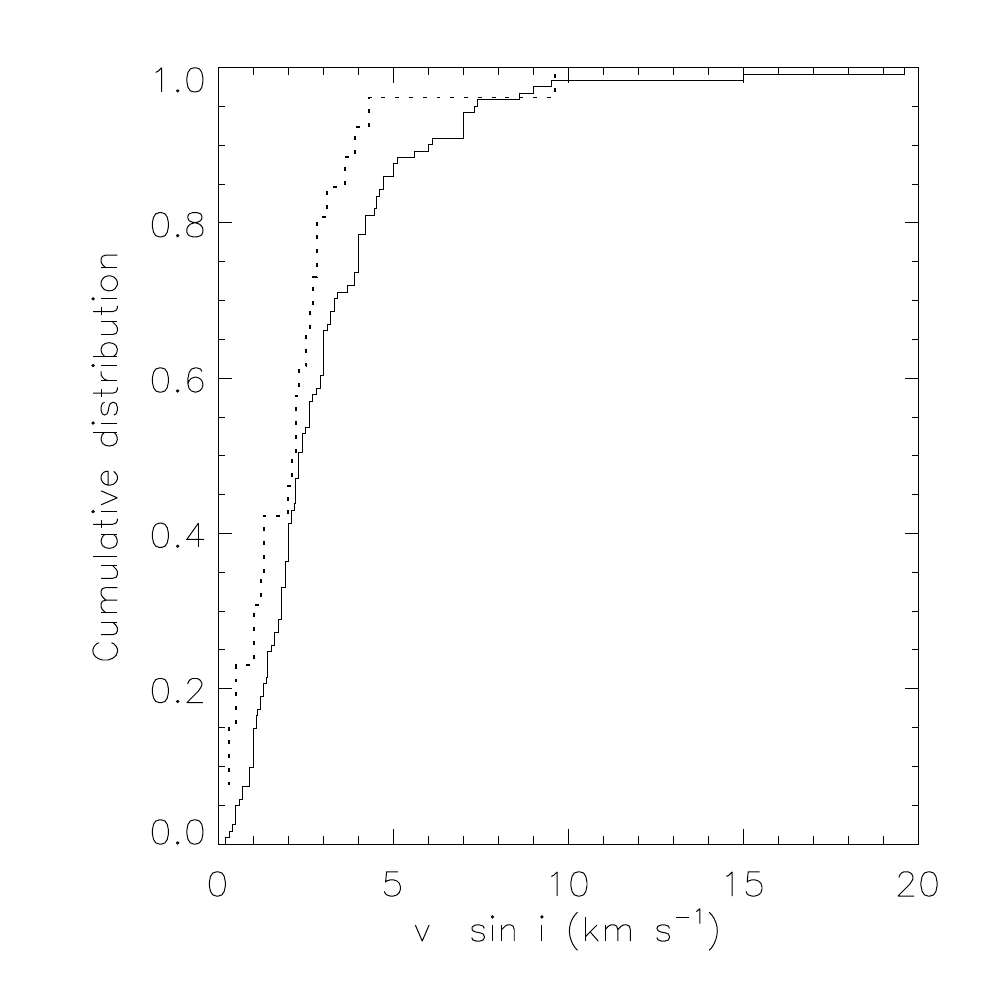}}
\caption[]{The cumulative distributions of $v \sin i$ for main–sequence stars with one planet (solid curve) and two or more planets (dashed curve). The analyzed samples of stars with one planet and two or more planets are composed, respectively, of 122 and 25 stars, the  latter hosting 62 planets.}
\label{fig5}
\end{figure}

To further explore a possible relationship between the rotation of stars and the presence of planets, we also applied the Kolmogorov-Smirnov test for $v \sin i$ distribution of stars with one, two or more planets, looking for a possible reflex of the number of planetary companions in the rotation of their parent stars. Fig. ~5 shows the cumulative $v \sin i$ functions for stars with one and two or more planets. In the present case, a probability value of 0.574 was obtained, indicating that the $v \sin i$ distributions for stars with one and two or more planets are consistent with samples being drawn from the same population distribution. However, these results should be interpreted cautiously, since they may be a result of  biases associated with $v \sin i$ measurement uncertainties or selection effects in the definition  of the two stellar samples, including stellar ages and masses. Accordingly, we  performed an additional K-S test to check the statistical nature of the distributions of effective temperature and age of the stars with and without detected planets. Fig.~6 shows the cumulative functions for the distributions of effective temperature (upper panel) and age (lower panel), for both the sample of F- and G-type main-sequence stars. Probability values of 0.268 and 0.236 were  obtained with  the K-S test for effective temperature and age, respectively, which is consistent with the present samples of stars with and without planets coming from populations with identical distributions in the HR Diagram. The same result was  obtained  with the  K-S test of the distributions of luminosity for the aforementioned  samples, with a probability value of 0.121.  { {It must be remembered that  stellar ages from isochrones have an uncertainty that varies dramatically with location in the HR Diagram, a fact that may  considerably compromise the K-S test for ages, in spite of the fact that we are considering only F and G-type main sequence stars.  Nevertheless, the most relevant point in our comparative analyses is the K-S test for Teff and luminosity, which presents  realistic uncertainties. }}
\begin{figure}
\centering
          \resizebox{1\hsize}{!}{\includegraphics[width=\linewidth]{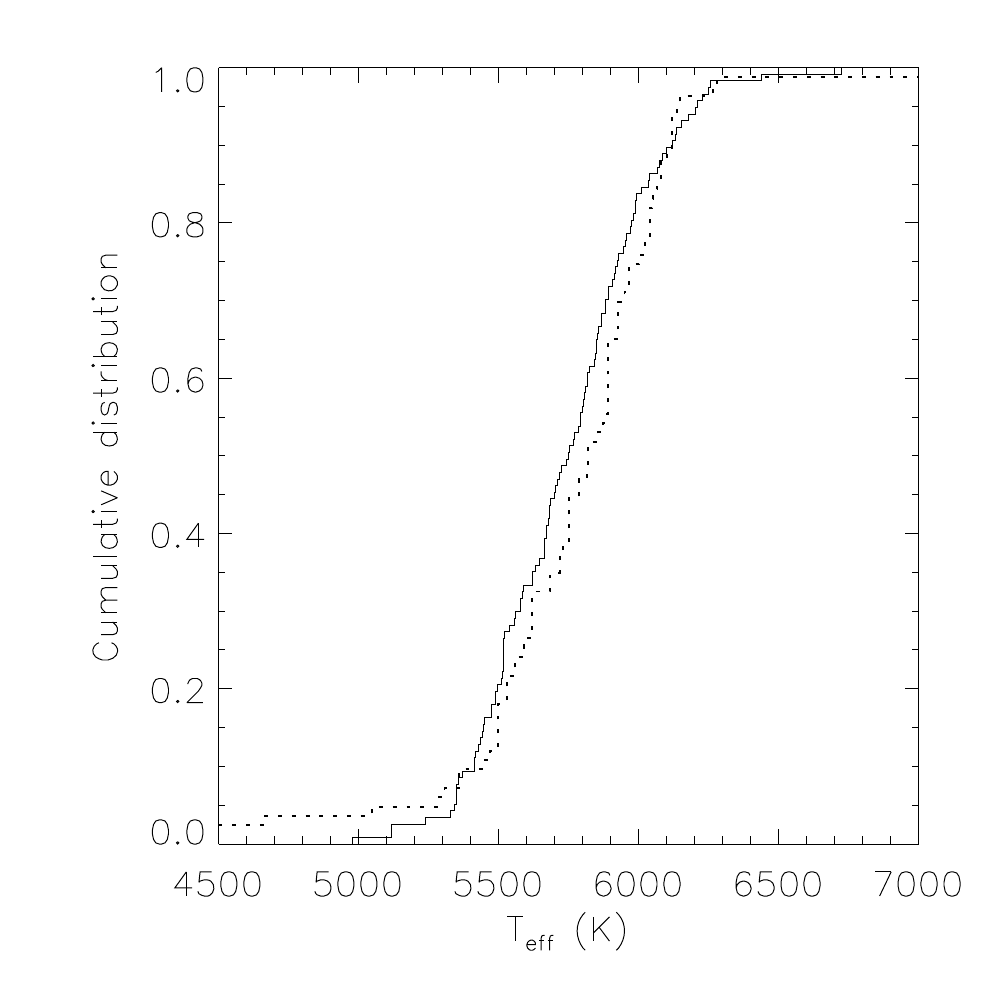}}\\
          \resizebox{1\hsize}{!}{\includegraphics[width=\linewidth]{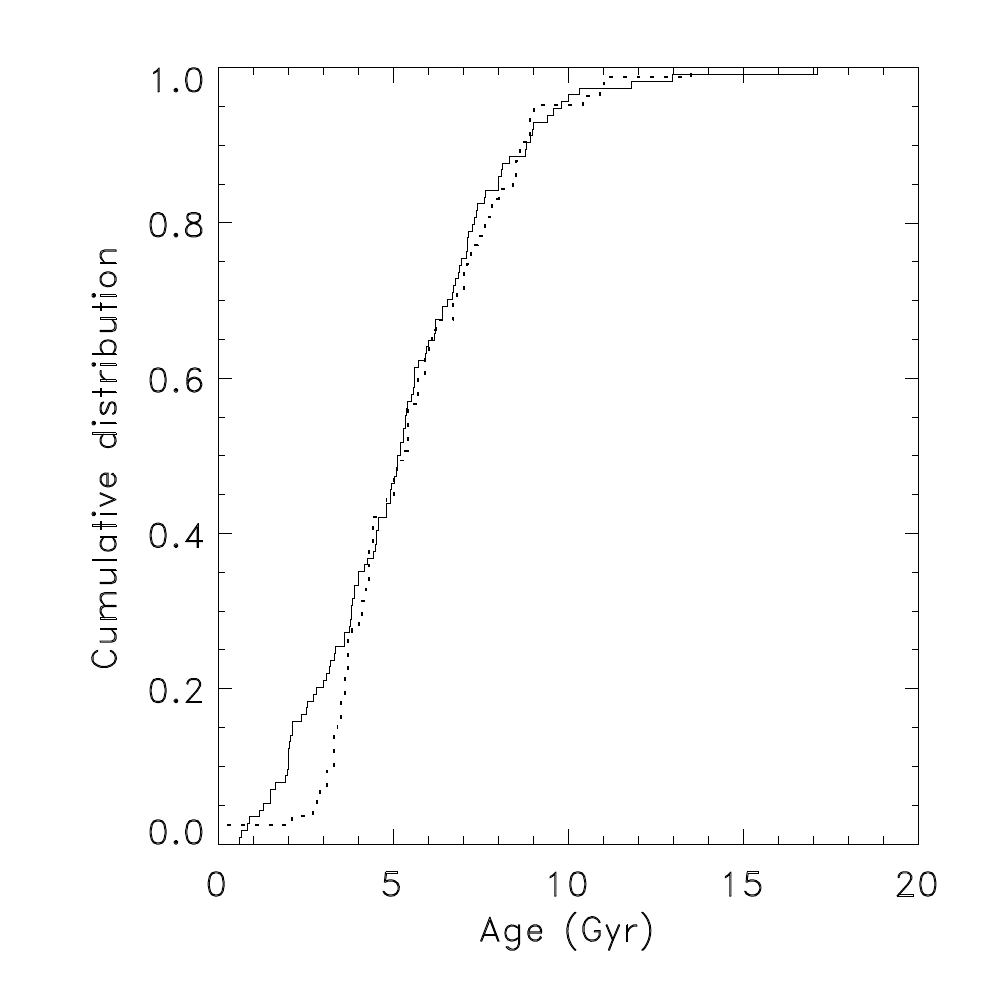}}\\
\caption[]{Comparison of the distributions of eﬀective temperature and age of F– and G–type main–sequence stars with and without planets. Solid and dashed curves stand for the cumulative functions for stars with and without planets, respectively.}
\label{fig6}
\end{figure}
\begin{figure}
\centering
\resizebox{1\hsize}{!}{\includegraphics{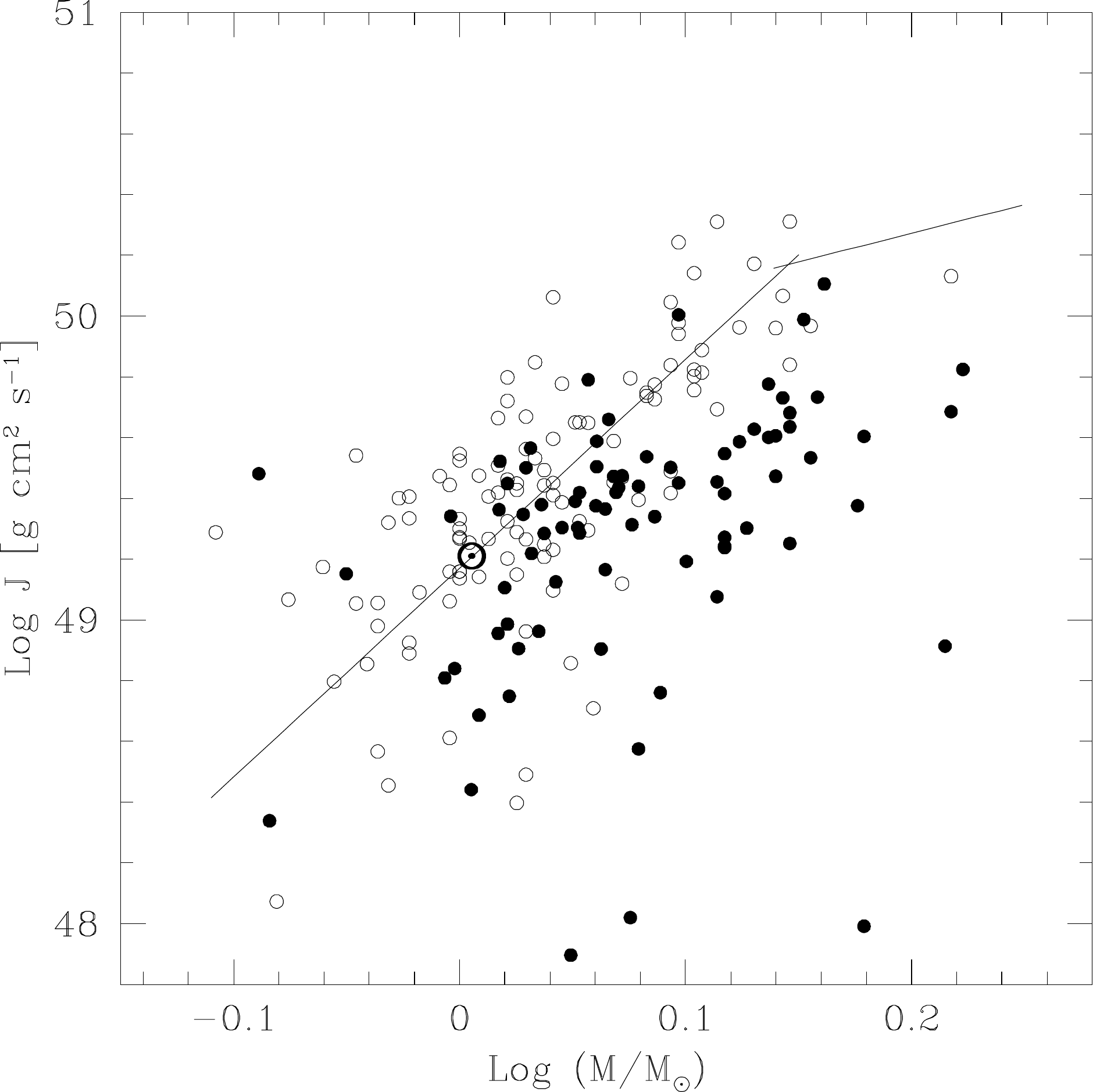}}
\caption[]{Stellar angular momentum versus stellar mass (in solar mass unity) for F– and G–type dwarf stars, corresponding to 118 stars with planets from the present working sample and 82 stars without detected planets from Bond (2008). In this ﬁgure, open and solid circles stand for stars with and without planets, respectively. The sun is also represented for comparative purposes. The solid line represents the best ﬁt of Kraft’s law (Kraft 1967; Kawaler 1987) $ J \propto (M/M_{\odot})^\alpha$, applied to the sample of stars with planets.}
\label{Fig7}
\end{figure}
\begin{figure}
\centering
          \resizebox{1\hsize}{!}{\includegraphics[width=\linewidth]{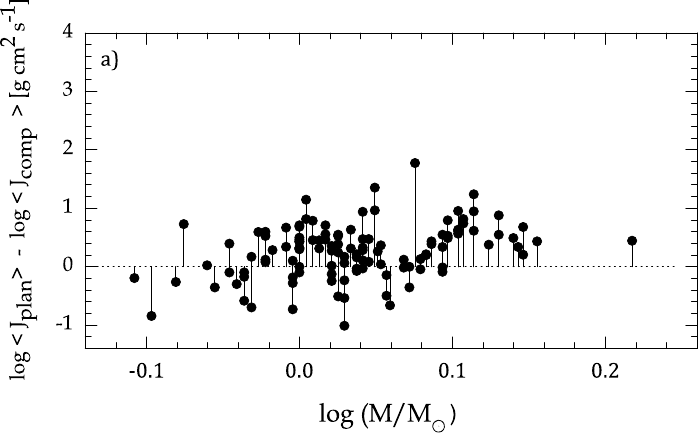}}\\
          \resizebox{1\hsize}{!}{\includegraphics[width=\linewidth]{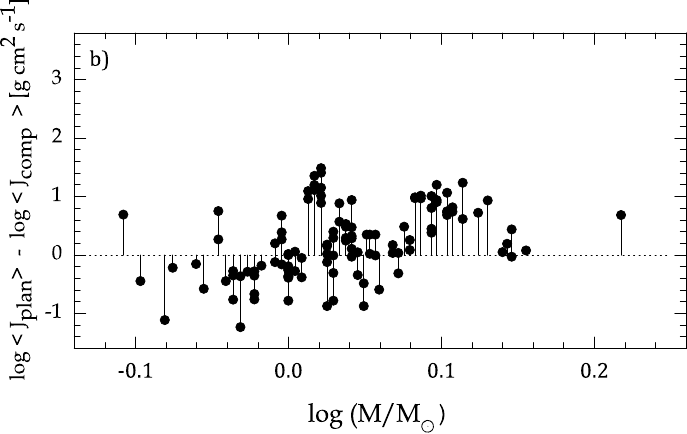}}\\
\caption[]{The angular momentum differences ($\log< J_{plan}>- \log <J_{comp}>$) between stars with planets $<J_{plan}>$ and comparison stars $<J_{comp}>$ represented in Fig.~7. Panel (a) shows the differences between stars with planets and the most similar comparison stars, from the stellar samples represented in Fig.~7, whereas the panels (b) correspond to stars with and without planets from the Geneva planet search survey.}
\label{Fig8}
\end{figure}

\subsection{The angular momentum of stars with planets}

One notable characteristic of the Solar System, with direct implications for its formation process, is the fact that most of its angular momentum is in planetary orbital motion.  For instance, total orbital angular momentum of the Solar System, largely due to Jupiter, is at least two orders of magnitude larger than the spin angular momentum of the Sun. By extending the study of angular momentum behavior to other stars with planets, we can examine how different or similar their spins are compared to the Sun.  In the following, we speculate very briefly on the angular momentum characteristics of the parent stars of extrasolar planets, by comparing the angular momentum of stars with and without detected planets. Further, we analyze the angular distribution for the two samples of stars in the context of Kraft’s well known law $ J \propto (M/M_{\odot})^\alpha\ $ (Kraft 1967; Kawaler 1987). 

For this purpose, we estimated the angular momentum for the F- and G-type dwarf stars of the present working samples, stars with and without planets, according to the recipe  described in Sect. 2a.  { {Let us recall that, for the computation of the angular momentum we  used masses, determined spectroscopically by  Schneider (2009) and Valenti and Fischer (2005),  for stars with and without planets, respectively. Indeed, as pointed out by Valenti and Fischer (2005) their  spectroscopically computed masses have a median fractional precision of 15\% and are systematically 10\% higher than masses from interpolating isochrones. Such a fact explains the presence of a dozen  stars  in our analyses with M $\gtrsim$ 1.25~M$_{\odot}$, in contrast to the distribution of stars without planets in the HR Diagram displayed in the previous section.}} Fig.~7 shows the distribution of the angular momentum of the main-sequence F- and G-type stars that make up  our working sample, as a function of  stellar mass (in solar mass unity). In this figure, open and solid symbols stand for stars with and without planets, respectively. The solid line represents the best fit of Kraft’s law $ J \propto (M/M_{\odot})^\alpha$, applied to the sample of stars with planets, with the exponent $\alpha$ of the power law remaining as a free parameter. It is immediately apparent that the relation found by Kraft applies  to  the parent stars of extrasolar planets.  Another interesting aspect emerges from Fig.~7, when  we compare the angular momentum distribution of the parent stars of planets with that of stars without detected planets. In spite of the more limited number of stars without planets, they show a clear trend of having an angular momentum in deficit,  compared to the stars with planets.  In particular for masses larger than about 1.25~M$_{\odot}$, stars without detected planets tend to have a lower mean angular momentum than stars hosting planets, with the majority of stars below Kraft’s law. Indeed, the same scenario is observed when  we analyze the angular momentum distribution of similar samples of main-sequence F- and G-type stars with and without planets, listed in the Geneva and AAPS search planet surveys.  { These trends remain unchanged  if  we take masses from Valenti and Fischer (2005), obtained by interpolating isochrones, for the computation of the angular momentum of stars without planets.} 

As a check of the patterns observed in Fig.~7, we applied the procedure proposed by Gonzalez (2008) for comparison of stellar properties of stars with and without planets, which estimate a measure of the proximity of two stars in [Teff, log g, [Fe/H], Mv] space. Panel (a) of Fig.~8 shows the angular momentum differences ($\log<J_{plan} >-\log<J_{comp}>$) between stars with planets $<J_{comp}>$ and the most similar comparison stars $<J_{comp}>$, taking into account the stars of the two samples represented in Fig.~7. The evident trend towards an angular momentum deficit in stars without planets, with masses larger than about 1.25~M$_{\odot}$, shown in Fig.~7, remains unchanged in this new analysis. Again, the same scenario is observed when  we consider similar samples of main-sequence F- and G-type stars with and without planets listed in the Geneva search planet survey, as illustrated in panels (b) of Fig.~8. One important aspect in favor of the apparent discrepancy observed in the distribution of the angular momentum of stars with planets versus the one for stars without detected planets, with an angular momentum deficit in stars without planets, is the fact that both samples appear to be statistically indistinguishable, as shown in the previous section. 

\begin{figure*}
\centering
       \begin{minipage}[b]{0.33 \linewidth}
          \resizebox{1\hsize}{!}{\includegraphics[width=\linewidth]{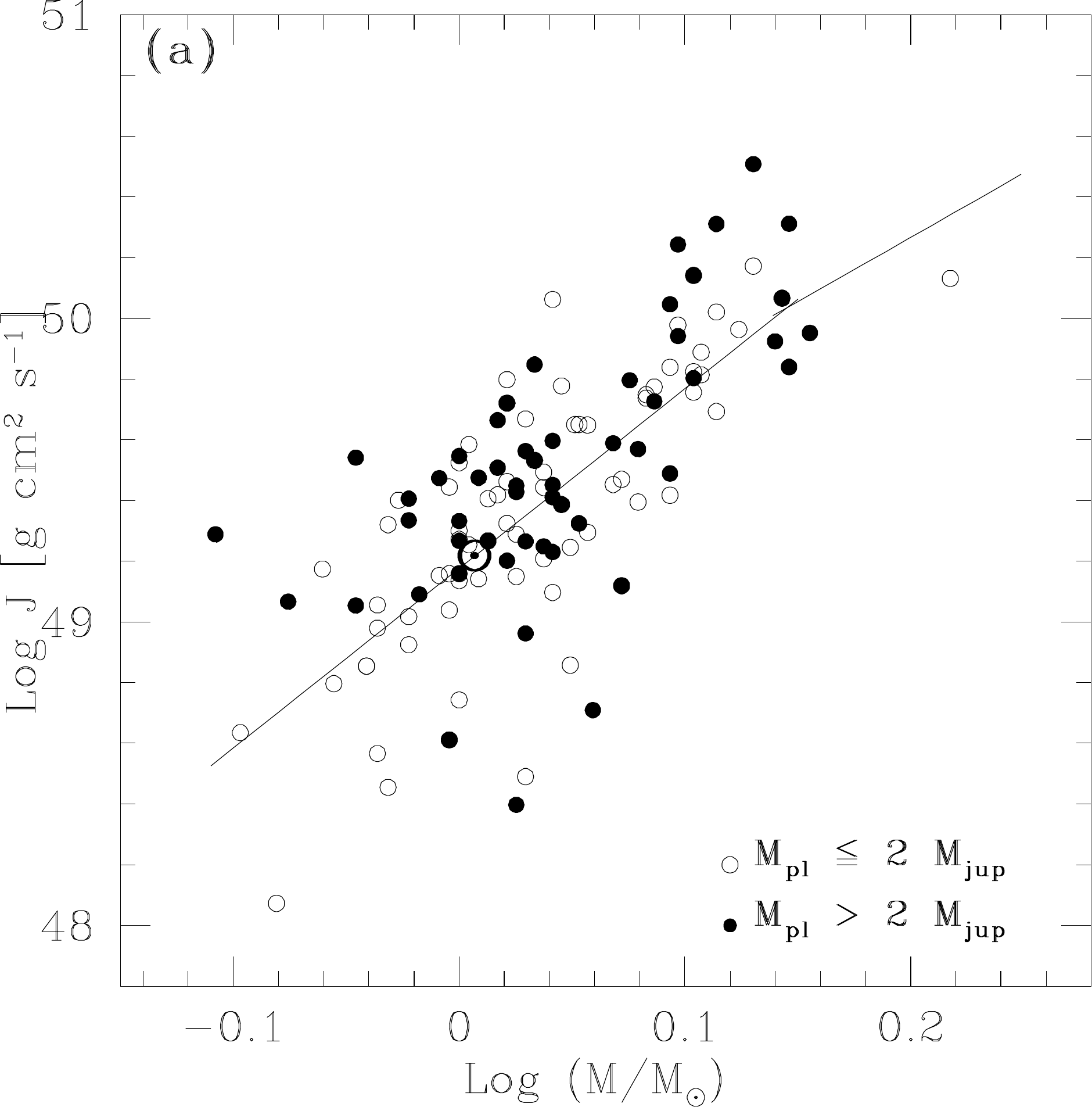}}\\
       \end{minipage}\hfill
       \begin{minipage}[b]{0.33 \linewidth}
          \resizebox{1\hsize}{!}{\includegraphics[width=\linewidth]{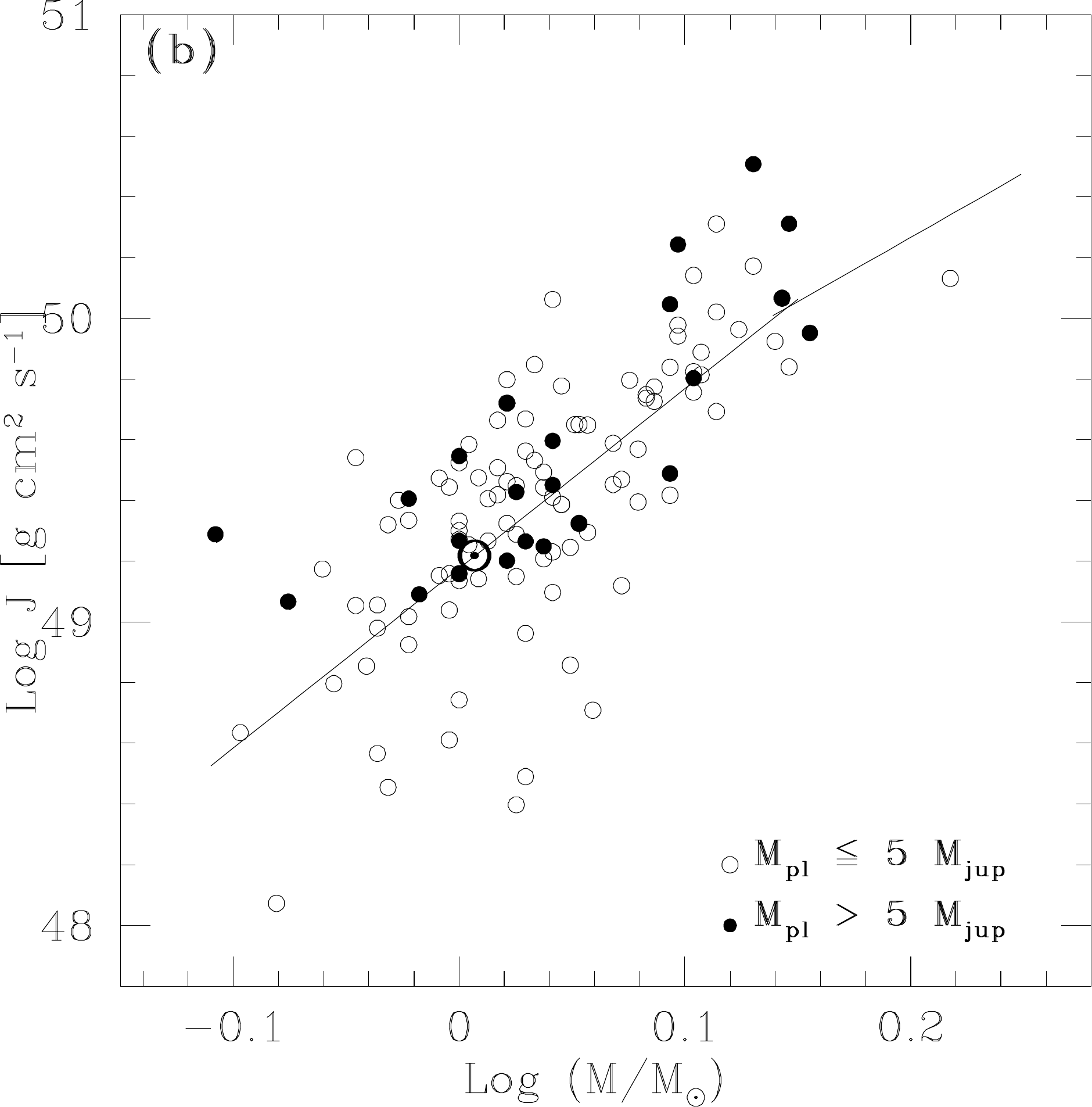}}\\
       \end{minipage}\hfill
       \begin{minipage}[b]{0.33 \linewidth}
           \resizebox{1\hsize}{!}{\includegraphics[width=\linewidth]{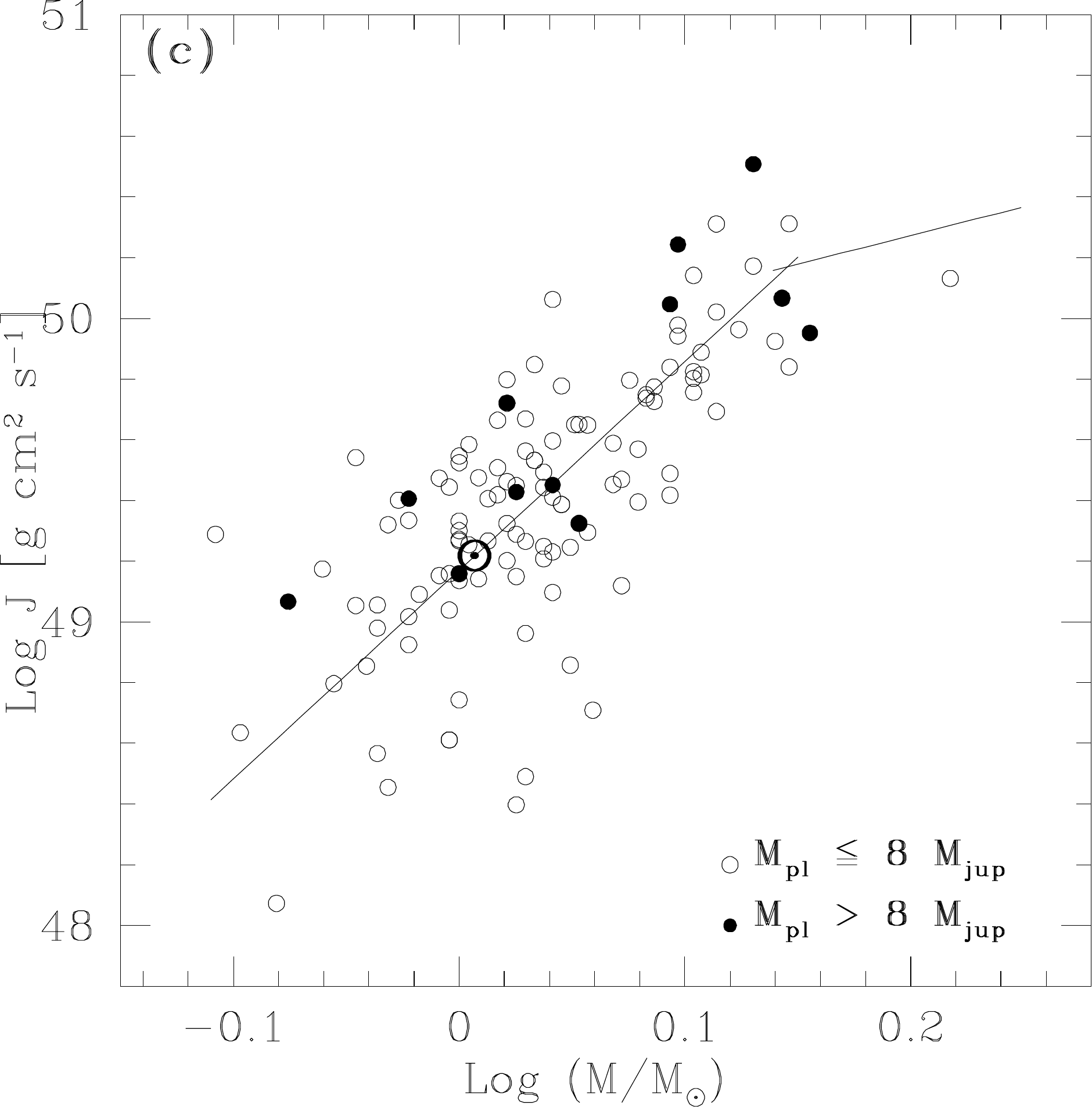}}\\
       \end{minipage}
\caption[]{Stellar angular momentum versus stellar mass (in solar mass unity) for the 118 F– and G– type dwarf stars with planets of the present sample, with the stars segregated by the mass strength of their orbiting planets. The planetary mass intervals considered are indicated in the box. The sun is also represented for comparative purposes. As in Fig. 8, the solid line represents the best ﬁt of Kraft’s law $ J \propto (M/M_{\odot})^\alpha$ (Kraft 1967; Kawaler 1987), applied to the sample of stars with planets.} 
\label{Fig9a; Fig9b; Fig9c}
\end{figure*}
We turn now to an analysis of the role of planetary mass on the angular momentum $J(M)$ of the parent stars. Fig.~9 shows the stellar angular momentum distribution of 118 F- and G-type dwarf stars with planets, now separated into three mass intervals of their orbiting planets, namely $2.0 M_{\rm Jup} < M_{\rm pl}~\leq ~2.0 M_{\rm Jup}$, $5.0 M_{\rm Jup} < M_{\rm pl} \leq 5.0 M_{\rm Jup}$ and $8.0 M_{\rm Jup} < M_{\rm pl} \leq 8.0 M_{\rm Jup}$, represented in panels (a), (b) and (c), respectively. A close inspection of these panels shows that stars hosting more massive planets tend to have the highest angular momentum, whereas stars with the least  massive planets tend to have the lowest. Interestingly, stars hosting planets with the largest mass tend to have angular momentum larger than the Sun. 
\section{Conclusions}

The present work conducts an unprecedented analysis of the rotational velocity behavior of the parent stars of extrasolar planets. We combined the projected rotational velocity $v \sin i$ of stars with planets with stellar and planetary parameters, for a unique sample of 147 stars, amounting to 184 extrasolar planets discovered by the radial-velocity procedure, including 25 multiple systems, searching for trends or anomalies in their relations.  In addition, for comparative purposes, we used a sample of stars without detected planets composed of 85 dwarf stars from the Anglo-Australian planet survey. Different Kolmogorov-Smirnov tests, performed with effective temperature, luminosity and age, indicate that both samples of stars come from populations with similar distributions in the HR Diagram. 

First, we placed stars in the HR Diagram, from which one observes that stars with planets follow the well established scenario for the rotation of intermediate to low main-sequence stars, first pointed out by Kraft (1967), with a sudden decline in rotation near 1.2~\Msun. Stars with planets exhibiting enhanced rotation are mostly those with effective temperature higher than about 6000~K and luminosity higher than the solar value, whereas slow rotators are mostly the coolest stars, with luminosity lower than the solar value.  This transition, fast rotators and moderate to slow rotators, is explained as resulting from the presence of a convective envelope in late-type stars. This convective envelope slows down the stars from an initially rapid rotation rate through magnetic breaking, associated to the fact that these stars, on average, are older than the early-type and have undergone a longer  spin–down period. Qualitatively, stars without detected planets appear to follow the same scenario. The distributions of $v \sin i$ for stars with and without detected planets suggest that they are  being drawn from the same parent population.

We also analyzed the stellar angular momentum behavior as a function of stellar mass, in which one observes that stars with planets follow, at least qualitatively,  Kraft’s law J$ J(M) \propto (M/M_{\odot})^\alpha$. Further, we  compared the distribution of angular momentum $J(M)$ in  parent stars of extrasolar planets versus that of stars without detected planets.  The latter shows a clear trend of being in angular momentum deficit compared to the stars with planets. Furthermore, we observed that the vast majority of stars without detected planets are also in angular momentum deficit compared to the Sun.  If these tentative conclusions are  confirmed, it would imply that the angular momentum distribution in these two families of stars followed different histories, that is, stars with detected planets following  the solar system history. Finally, we  analyzed the distribution of the angular momentum of stars with planets as a function of planetary mass. Here, stars with more massive planets tend to have the highest angular momentum in relation to the solar value. 

As we underscored previously, the sample of stars hosting planets is subject to a number of  subtle biases, some showing the effect of masking the rotational behavior of such stars, including the relationship between stellar rotation and planetary orbital parameters.  Perhaps the major bias in the present study arises from the difficulty of detecting planets around fast rotators, since, to date, the main planet detection process, namely the Doppler method, cannot account for stellar spectra with broad lines. A large $v \sin i$ value results in broad lines and is a clear source of noise in Doppler measurements (Wright 2005). Again, planetary mass may also represent an additional observational bias in the study of the link between rotation and stellar and planetary parameters, in particular because the actual masses of the vast majority of extrasolar planets are unknown.  In addition, as already highlighted, all studies on planetary mass show a strong observational selection effect stemming  from the difficulty of detecting low-mass long-period planets, due to their small radial velocity amplitudes. Despite these limitations, the present work points to some interesting trends, which deserve a more detailed analysis to draw firmer conclusions, including a much larger detection sample , particularly of stars with low mass planets and large orbital periods.

The most interesting finding in the present work is undoubtedly  the discrepancy in the  the angular momentum behavior of stars with and without planets, with the majority of stars without detected planet exhibiting  a deficit in angular momentum compared to  hosting planet stars. If stars with  as yet  undetected planets  are also hosting planetary systems, their planets should have  different masses and orbital parameters from those already detected. Such an aspect may have a strong impact on the  angular momentum behavior and structure of these stars. For instance, one can expect that the total angular momentum of these as yet undetected planetary systems is more distributed among the planets, an aspect reflected by a deficit in the angular momentum of the star compared  to  present day stars with detected planets. 

The study of the rotation behavior in a larger sample of stars with planets, detected by the transit procedure, may help us  clarify some of the trends observed in this study involving planetary mass. Infrared modulation in planets orbiting stars with enhanced rotation could also reveal a signature of pseudosynchronized planetary rotational periods. Asteroseismological studies of stars with and without planets may also help us  elucidate the discrepancy in the angular momentum of these two families of stars, revealing peculiarities in the inner stellar structure with direct impact on the angular momentum evolution, such as  the degree of differential rotation.  The different patterns observed during this work could also have been influenced  by important biases associated to the small sample of stars without planets and different precisions in the $v \sin i$ measurements of both stellar samples. An essential first step is to determine the $v \sin i$ measurements of both stellar samples using the same procedure.

\section*{Acknowledgments}
This work was supported by grants from the Brazilian Agency CNPq. S. Alves thanks the Brazilian Agency CAPES for a graduate student fellowship.  We are also grateful to the Rio Grande do Norte Agency FAPERN for financial support. We thank the anonymous referee for suggestions and criticisms that greatly improved this paper as well as I. C. Le\~ao for his kind support with statistical analysis.

\bsp
\label{lastpage}
\end{document}